\documentclass[12pt]{article} 
\usepackage{jheppub}
\usepackage{graphicx, amsmath, xcolor, adjustbox}

\usepackage{bm}        
\usepackage{amsfonts}  
\usepackage{amsmath}   
\usepackage{amssymb}
\usepackage{slashed}

\usepackage{tikz}
\usepackage{tikz-feynman}
\usepackage{subfig}
\makeatother
\usepackage{multirow}
\usepackage{fancyhdr}
\usepackage{longtable}
\usepackage{parskip}
\usepackage{dcolumn}   

\def\equationautorefname~#1\null{eq.\,(#1)\null}

\makeatletter\g@addto@macro\bfseries{\boldmath}\makeatother

\DeclareRobustCommand\hbar{{\mathchar'26\mkern-9muh}}

\newcommand{\bea}{\begin{eqnarray}}
\newcommand{\eea}{\end{eqnarray}}
\newcommand{\beq}{\begin{equation}}
\newcommand{\eeq}{\end{equation}}

\usepackage[normalem]{ulem}

\def\CERN{CERN, Theoretical Physics Department, Geneva, Switzerland}
\def\Cambridge{DAMTP, University of Cambridge, Wilberforce Road, Cambridge, UK}
\def\TAMU{Department of Physics and Astronomy, Mitchell Institute for Fundamental Physics and Astronomy, Texas A\&M University, College Station, TX 77843, USA}

\begin{document}

\title{How Broad is a Neutrino?}

\author[a,b]{Hannah Banks,}
\author[b,c]{Kevin J. Kelly,}
\author[b]{and Matthew McCullough}
\affiliation[a]{\Cambridge}
\affiliation[b]{\CERN}
\affiliation[c]{\TAMU}

\emailAdd{hmb61@cam.ac.uk}
\emailAdd{kjkelly@tamu.edu}
\emailAdd{matthew.mccullough@cern.ch}
\date{\today}

\abstract{
Canonical neutrino oscillations arise due to the propagation of three mass eigenstates from production to detection.  We aspire to capture, in one simple framework, a broad range of new physics effects on neutrino propagation beyond this canonical picture -- this can be done by promoting the neutrino propagators to the general K\"all\'en-Lehmann form.  In this work we demonstrate how models predicting additional light propagating species of neutrino are naturally accommodated in this language and propose a simple model spectrum composed of just three `broadened' states as a flexible ansatz by which to explore the phenomenology of new physics in neutrino propagation. Reinterpreting existing neutrino oscillation measurements, we illustrate how this framework provides the capacity to probe deviations from the standard three-neutrino scenario systematically and generally. Whilst current data allows for relatively strong constraints on broadened neutrinos, we find the upcoming JUNO experiment will yield significant improvements, particularly for the heaviest neutrino, paving the way to a clearer understanding of how neutrinos propagate in vacuum.
}

\preprint{CERN-TH-2022-152}

\maketitle

\section{Introduction} 
Since their discovery, neutrinos have been an invaluable probe of physics `beyond the Standard Model' (BSM). Indeed the observation of flavour changes in neutrino oscillations, which require non-zero neutrino masses, is arguably one of the strongest pieces of evidence for new BSM physics.  This observation behooves particle physicists to probe, to the greatest degree possible, all aspects of the neutrino sector of the SM.

As neutrino oscillations arise as a result of massive neutrino propagation from the point of production to the point of detection, one may ask if these massive fermions are propagating from one point to another as expected.  In this work we aim to develop a coherent framework in which this question may be posed theoretically and answered experimentally.\footnote{Note that very recently Ref.~\cite{Gherghetta:2022ynp} appeared, with broadly similar motivations although very different specific considerations.}

\begin{figure}[t]
\centering
\includegraphics[width=0.4\textwidth]{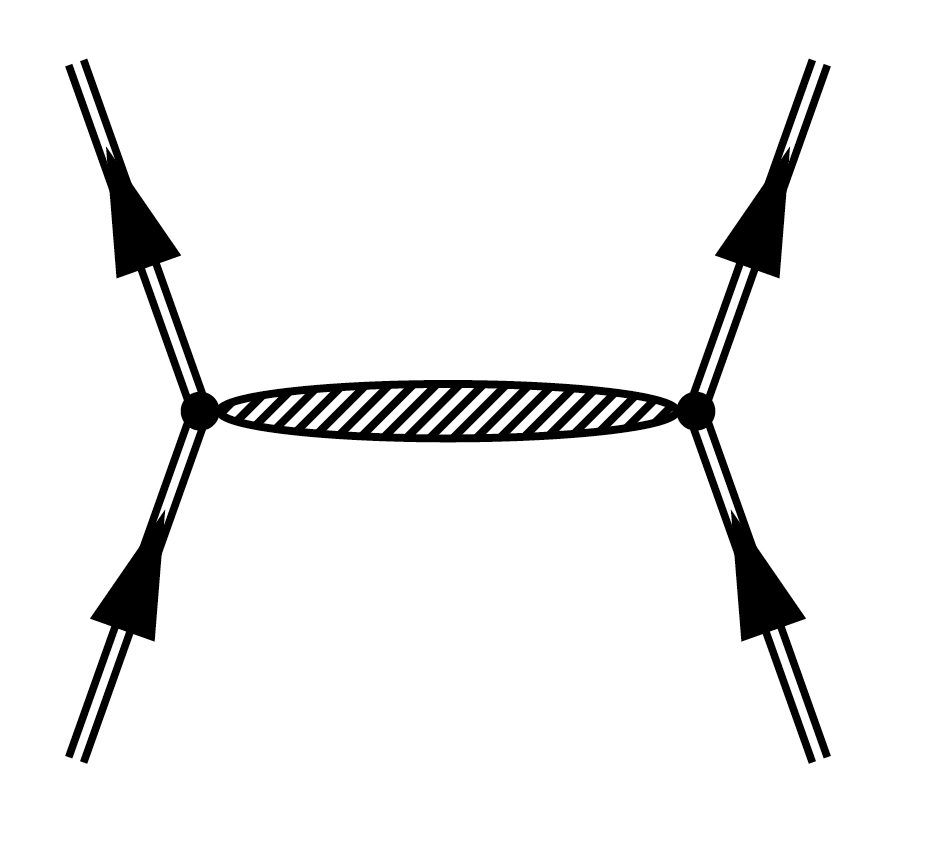}
\caption{A schematic Feynman diagram for a general neutrino propagation process. The double lines show the external states at the production and detection vertices. The elliptical blob symbolises the K\"all\'en-Lehmann propagator. In the case of canonical oscillations, this reduces to the usual Feynman propagator. }
\label{fig:KL}
\end{figure}

In any free or interacting quantum field theory the propagator (two-point function) for a fermion may be captured by a K\"all\'en-Lehmann representation.  Note that this is a non-perturbative representation, not reliant on any perturbative expansion, but only on the basic axioms of quantum mechanics.  As a result, this representation may be included in any phenomenological description of neutrino oscillations, whether QM-like or QFT-like, and it will capture any QFT-compatible BSM modifications of neutrino propagation from production to detection, as illustrated schematically in Fig.~\ref{fig:KL}.    To this end, in Section~\ref{sec:KLTheory} we generalise the usual neutrino oscillation framework to include the more general K\"all\'en-Lehmann form, propagating this general form all the way through to generalised formul{\ae} for oscillation probabilities of neutrino appearance and disappearance.

To illustrate the utility of this approach more concretely we provide two simple BSM scenarios in which the neutrino propagator becomes `broad'.  The first is the previously studied `Pseudo-Dirac' neutrino model \cite{Wolfenstein:1981kw,Petcov:1982ya,Bilenky:1983wt,Kobayashi:2000md,deGouvea:2009fp,Anamiati:2017rxw,Anamiati:2019maf,Martinez-Soler:2021unz}, where neutrino masses are effectively Dirac and a small amount of lepton number violation leads to a small splitting of mass eigenstates.  A second model generalises this setup further, through a fermionic clockwork-inspired deconstruction of fermions on a circle, to give a broad band of states in lieu of a single neutrino mass eigenstate.  The mapping of both models into the K\"all\'en-Lehmann representation of the propagator is developed and their impact on observable neutrino oscillation data is investigated.  Motivated by these models and the fact that an almost limitless zoology of models is in principle realisable, we then present a simple `top-hat' phenomenological ansatz which captures the dominant oscillation features of more complete microscopic models that lead to band-like neutrino spectral functions.

Armed with this formalism and utilitarian phenomenological ansatz, we first discuss how observations of neutrinos from distant sources can aide in testing these scenarios in Section~\ref{sec:Decoherence}. Then, in Section~\ref{sec:Experiments} we take the next step to see how well we currently understand BSM effects in neutrino propagation as it pertains to Earth-based experiments.  In practical terms we do this by determining how well KamLAND, Daya Bay, T2K and NOvA measurements constrain non-SM contributions to neutrino propagation, finding that KamLAND has been particularly powerful.  On the other hand, the upcoming JUNO experiment will provide an even more powerful and complementary probe, breaking flat directions that presently exist.  Finally, Section~\ref{sec:Conclusions} offers some concluding remarks.

\section{Neutrino Propagation from K\"all\'en-Lehmann}\label{sec:KLTheory}

There is a broad literature concerning neutrino oscillation amplitudes, ranging from textbook quantum-mechanical derivations to involved quantum field theory treatments.  There is little to be gained in repeating these analyses here, thus we focus on the core novel ingredient of this work, wherein the key addition to typical treatments, reviewed for example in \cite{Beuthe:2001rc,Kopp}, is to replace the free neutrino propagator and matrix elements found in QFT treatments
\beq
 \sum_j U^*_{\alpha j} U_{\beta j} \frac{\slashed p+m_j}{p^2-m_j^2+i \epsilon} ~~,
\eeq
by the more general K\"all\'en-Lehmann propagator for spin-1/2 particles
\beq
\label{eq:KL}
G^{\text{KL}}_{\alpha \beta}(p^2) = i \int_0^{\infty} d\mu^2  \frac{\slashed p\tilde{\rho}_{\alpha \beta}(\mu^2)+\rho_{\alpha \beta}(\mu^2)}{p^2 - \mu^2 + i\epsilon} ~~,
\eeq
 where $\alpha$ and $\beta$ denote the flavour eigenstate at the point of production and detection respectively.
This captures any new-physics effects on the propagation of neutrinos consistent with the axioms of QFT and, essentially, promotes the sum over mass eigenstates to a continuum integral. The $\rho$ functions here encode both the density of states, and the matrix elements describing the overlap between the mass and interaction eigenstates. One recovers the usual Feynman propagator if $\tilde{\rho}(\mu^2)$ and $\rho(\mu^2)/\mu$  are identical and comprise delta functions $\delta(\mu^2-m_j^2)$. More general scenarios such as neutrino mixing with hidden sector states, including for strongly-coupled hidden sectors, can be described by functions exhibiting a discrete or continuous density of states.  

Since the propagator above and the spectral functions $\rho_{\alpha\beta}(\mu^2)$ and $\tilde{\rho}_{\alpha\beta}(\mu^2)$ are expressed in the interaction basis, whose eigenstates need not coincide with those of the vacuum Hamiltonion, the spectral functions need not be real nor obey the usual positivity requirements satisfied by K\"all\'en-Lehmann functions.
An exception is in the case $\alpha=\beta$, when the mapping between mass- and interaction-bases depends only on real, positive factors such as $|U_{\alpha j}|^2$. In this case, the spectral functions are required to be real, and both $\tilde{\rho}_{\alpha\alpha}(\mu^2)$ and $(\mu \tilde{\rho}_{\alpha\alpha}(\mu^2) - \rho_{\alpha\alpha}(\mu^2) )$ are positive-definite.

In all scenarios of interest,  the neutrinos are ultra-relativistic. In this limit,  chirality and helicity eigenstates coincide and the spin structure can be factored out of the two-point function. All  the relevant oscillation physics is fully captured by the scalar amplitude which itself can be represented via the scalar K\"all\'en-Lehmann propagator. In practice this amounts to setting  $\tilde{\rho}(\mu^2) = 0$  in Eq.~\ref{eq:KL}.  Cataloguing the BSM possibilities for the neutrino sector thus fundamentally reduces to the study of a single scalar function,  $\rho(\mu^2)$,  which will be the object of interest in what follows. 

Following the standard QFT calculations with this modification, the probability for neutrino flavour transitions for propagating neutrinos of momentum $\bold{p}$ is given by
\beq
\label{prob}
P_{\alpha\beta} = \left | \int_0^{\infty} d\mu^2  e^{-i \sqrt{\mu^2 + \bold{p}^2}L}\rho_{\alpha\beta}(\mu^2) \right | ^2 ~~.
\eeq
In this statement we have applied the approximation $L\approx T$, and are explicitly working in the limit in which the full QFT calculation reproduces that obtained in QM. This requires the additional assumption that the neutrino at the production and detection vertices can be approximated as a plane wave, or equivalently that the neutrino wavepackets maintain coherence over the entire distance of interest, $L$. 

In order to expand $\sqrt{\mu^2 + \bold{p}^2}$, we must assume that all coherently-propagating neutrinos are relativistic. This requires that $\rho_{\alpha\beta}(\mu^2)$ does not have support for large $\mu^2$, which is expected for light, oscillating neutrinos with large (above ${\sim}$MeV) energies. We thus have that  $\sqrt{\mu^2 + \bold{p}^2} \approx E + \mu^2/(2E)$, and can express the transition probability as
\begin{equation}
P_{\alpha\beta} =  \left | \int_0^{\infty} d\mu^2  e^{-i \frac{\mu^2 L}{2E} }\rho_{\alpha\beta}(\mu^2) \right | ^2  ~~.
\end{equation}
This expression for the transition amplitude is simply the Fourier transform of the scalar spectral function. 

\subsection{Two-Flavour Example}
To make concrete headway we now turn to two-flavour mixing as characterised by the unitary matrix 
\begin{equation}
U = \begin{pmatrix}
\cos \theta & \sin \theta \\
-\sin \theta & \cos \theta
\end{pmatrix} ~~,
\end{equation}
which rotates flavour states $(\nu_\alpha, \nu_\beta)$ into mass eigenstates $(\nu_1, \nu_2)$.  For example, the $\nu_\alpha$ survival probability thus follows from the spectral density
\beq
\rho_{\alpha\alpha} =  \delta(\mu^2 - m^2_1) \cos^2 \theta- \delta(\mu^2- m^2_2)\sin^2 \theta ~~,
\eeq
giving
\begin{equation}
P\left(\nu_\alpha \to \nu_\alpha\right) = 1 - \sin^2 2\theta \sin^2 \left( \frac{\Delta m_{12}^2 L}{4E} \right) ~~,
\end{equation}
where $\Delta m_{21}^2 \equiv m_2^2 - m_1^2$. Similarly, the $\nu_\alpha \rightarrow \nu_\beta$ ($\alpha\neq\beta$) transition probability follows from
\beq
\rho_{\alpha\beta} =  \sin \theta \cos \theta \left(\delta(\mu^2 - m^2_1) - \delta(\mu^2- m^2_2) \right) ~~,
\eeq
giving
\begin{equation}
P\left(\nu_\alpha \to \nu_\beta\right) = \sin^2 2\theta \sin^2 \left( \frac{\Delta m_{12}^2 L}{4E} \right) ~~.
\end{equation}
We see that the predictions from this formalism map to textbook expressions in this standard case.

\subsection{Pseudo-Dirac Neutrinos}\label{subsec:PD}
A simple and well-known possibility that converts a single neutrino mass eigenstate into multiple states is that of pseudo-Dirac neutrinos \cite{Wolfenstein:1981kw,Petcov:1982ya,Bilenky:1983wt,Kobayashi:2000md,deGouvea:2009fp,Anamiati:2017rxw,Anamiati:2019maf,Martinez-Soler:2021unz}.  Structurally, the model consists of Dirac neutrinos, preserving a lepton-number symmetry, supplemented by a small source of explicit lepton-number violation which can be naturally small.   This symmetry breaking thus splits the Dirac neutrinos into two Majorana mass eigenstates.  Explicitly, the usual Lagrangian is
\beq
\lambda \nu_R L H + h.c.,
\eeq
which (after electroweak symmetry breaking) generates a Dirac mass $m_D = \lambda v/\sqrt{2}$.  The small lepton-number violating Majorana mass term for the right-handed neutrino is
\beq
\frac{1}{2} M \nu_R^2 +h.c.
\eeq
In the limit $m_D \gg M$, this has physical mass eigenvalues
\beq
M_{\pm}=m_D \pm \frac{M}{2} ~~,
\eeq
and the mixing angle between the EW gauge eigenstate and the mass eigenstates is
\beq
\tan (2 \theta) = \frac{2 m_D}{M} ~~.
\eeq
When $m_D \gg M$ the mixing angle is approximately maximal, $\theta \approx \pi/4$.  This furnishes a basic example where the mass eigenstate that propagates from production to detection is, microscopically, multiple states of different mass.  

This can be generalised to three flavours by the introduction of 3 right-handed neutrinos, $\nu_{Ri }$, for $i \in \{1,2,3\}$. For simplicity, we take the alignment limit where the $3 \times 3$ Dirac and Majorana mass matrices can be simultaneously diagonalised. We shall henceforth refer to the diagonal entries of these as $m_{D_i}$ and $M_{i}$ respectively. The $6 \times 6$ unitary matrix describing the rotation of the $\{\nu_{eL},\nu_{eR},\nu_{\mu L},\nu_{\mu R},\nu_{\tau L},\nu_{\tau R}\}$ flavour states into the mass eigenstates may be conveniently written as a product of $ 3 \times 3 $ and $2 \times 2$ matrices, defined as
\begin{equation}                                           
O_{\alpha \beta a b} = U_{\alpha \beta} R_{a b} (\theta_i)~~,
\end{equation}
where $U_{\alpha \beta}$ is the usual leptonic mixing matrix and $R_{ab}$ is a $2 \times 2$ rotation matrix with a rotation angle $\tan 2\theta_i = 2 m_{D_i}/M_{i}$.

The spectral density for $\alpha$ to $\beta$ flavour transitions is 
\begin{align}
\rho_{\alpha \beta} (\mu^2) =   \sum_{i = 1}^{3} U^*_{\alpha i} U_{\beta i} \left[ \cos^2(\theta_i)\delta(\mu^2 - M_{i,-}^2) + \sin^2(\theta_i)\delta(\mu^2 - M_{i,+}^2) \right] ~~,
\end{align}
where $M_{i,+}, M_{i,-}$ refer to the masses of the physical Majorana mass eigenstates into which the $i^{\textnormal{th}}$ Dirac neutrino splits.  Motivated by this, we now generalise further to a band of states.

\subsection{A Band of Neutrinos}\label{subsec:Band}
One way a band of neutrino states can be realised is through a `clockwork'-inspired~\cite{Choi:2015fiu,Kaplan:2015fuy,Giudice:2016yja} fermion ring.  Consider $N$ identical copies of the Standard Model enjoying a large translation symmetry in theory space.  These different sectors could have renormalisable (hence relevant at low energies) couplings to one another through one of three portals: Higgs, kinetic mixing and neutrino.  We will focus on the latter and suppose the following ring of couplings for a single flavour of neutrinos
\beq
\mathcal{L}_\lambda = \frac{\lambda}{q} \left[ \sum_{j=1}^{j=N-1} L_j H_j (\psi_j-q \psi_{j+1}) + L_N H_N (\psi_N-q \psi_{1}) \right] ~~,
\eeq
where the $\psi$ are SM-neutral Weyl fermions, $\lambda$ a small Yukawa coupling, and we assume $q>1$.  Upon electroweak symmetry breaking in all of the sectors the collective neutrino mass matrix, for one active flavour, becomes
\beq
M_\nu = \frac{m_D}{q}
\begin{pmatrix}
1 & -q & 0 & \cdots &  & 0 \cr
0 & 1 & -q & \cdots &  & 0 \cr
0 & 0 & 1 & \cdots & & 0 \cr
\vdots & \vdots & \vdots & \ddots & &\vdots \cr
 0 & 0 & 0 &\cdots & 1 & -q \cr
 -q & 0 & 0 &\cdots & 0 & 1
\end{pmatrix} \ ~~,
\label{psimass}
 \eeq
where $m_D = \lambda v/\sqrt{2}$.  The physical eigenvalues of this mass matrix are
\beq
m_j^2 = \frac{m_D^2}{q^2} \left(1+q^2 - 2 q \cos \left(\frac{2 \pi j}{N} \right) \right) ~~.
\eeq
We thus have a band of $N$ mass eigenstates centred at $\sim m_D$ with a breadth of $\sim m_D/q$.   The overlap between these states and the flavour eigenstates  is given by the elements of the rotation matrix
\beq
R_{jk} = \frac{\cos\left( \frac{2 \pi j k}{N} \right) + \sin\left( \frac{2 \pi j k}{N} \right)}{\sqrt{N }} ~~.
\eeq
Thus, for instance, the overlap between the $j^{\text{th}}$ mass eigenstate of a given generation and the interaction eigenstate in that sector of the ring is given by the elements $R_{1j}$. 

This can be further generalised to include 3 active generations of neutrino in each copy of the SM, with (site-independent) couplings $\lambda_{\alpha}$ where $\alpha \in \{1,2,3\}$. We take $q$ to be the same for each generation.  In our conventions, the spectral density describing  transitions from the $\alpha$ to $\beta$ flavour eigenstates is then
\begin{equation}
\rho_{\alpha \beta} = \sum_{i=1}^{3} U^*_{\alpha i }U_{\beta i}\left(\sum^{N}_{j= 1}  R_{1 j}^2\delta(\mu^2 - m_{ij}^2)\right) ~~,
\end{equation}
where the $m_{ij}$ refer to the masses of the $N$ physical mass eigenstates of the $i^{\textnormal{th}}$ generation.  This model essentially replaces any would-be SM Dirac neutrino mass eigenstate, of mass $\sim m_D$, by a band of states spread about this mass scale. 

Some comments are in order.  This model serves only to illustrate that such a scenario is possible, but is not intended to advertise the model as being particularly strongly motivated in its own right.  Furthermore, many details of the model are not necessary in order to realise the same qualitative scenario including, for instance, the translation symmetry that enforces equal $q$ at each site which could be softly broken.

Finally, a brief comment on cosmology.   If each sector were truly identical then it would be necessary that only the SM sector is reheated at the end of inflation, otherwise the neutrinos and photons of the hidden sectors would presumably lead to inconsistencies with cosmological observations.  It may also be the case that the reheating temperature is necessarily low to avoid over-populating the other neutrino states.

\subsection{Theory-Space Perspective}\label{subsec:TheorySpace}
To understand the phenomenology of these more exotic scenarios it is instructive to consider the process in terms of the `theory-space' sites, borrowing terminology from dimensional deconstruction \cite{Arkani-Hamed:2001kyx,Arkani-Hamed:2001nha}.  Consider an $N$-site band model.  Only one of the sites corresponds to an active, and potentially measurable neutrino. The remaining $N-1$ sites are sterile. When we talk of physically measurable neutrino oscillations, we refer to the probability of starting in the active flavour eigenstate of the ($\alpha$) generation and being measured in the active flavour eigenstate of the ($\beta$) generation at some later time, however here we will focus on a single-generation case.

Since the starting active flavour eigenstate is some superposition of the $N$ mass eigenstates through which the system evolves, there will be a generally non-zero probability of being in one of the other $N -1$ sites when a neutrino is detected.  In general, the higher the value of $N$, the lower the detection probabilities become due to the greater number of available sites to which the system has evolved  during propagation. 

\begin{figure}
\centering
\includegraphics[width=0.6\textwidth]{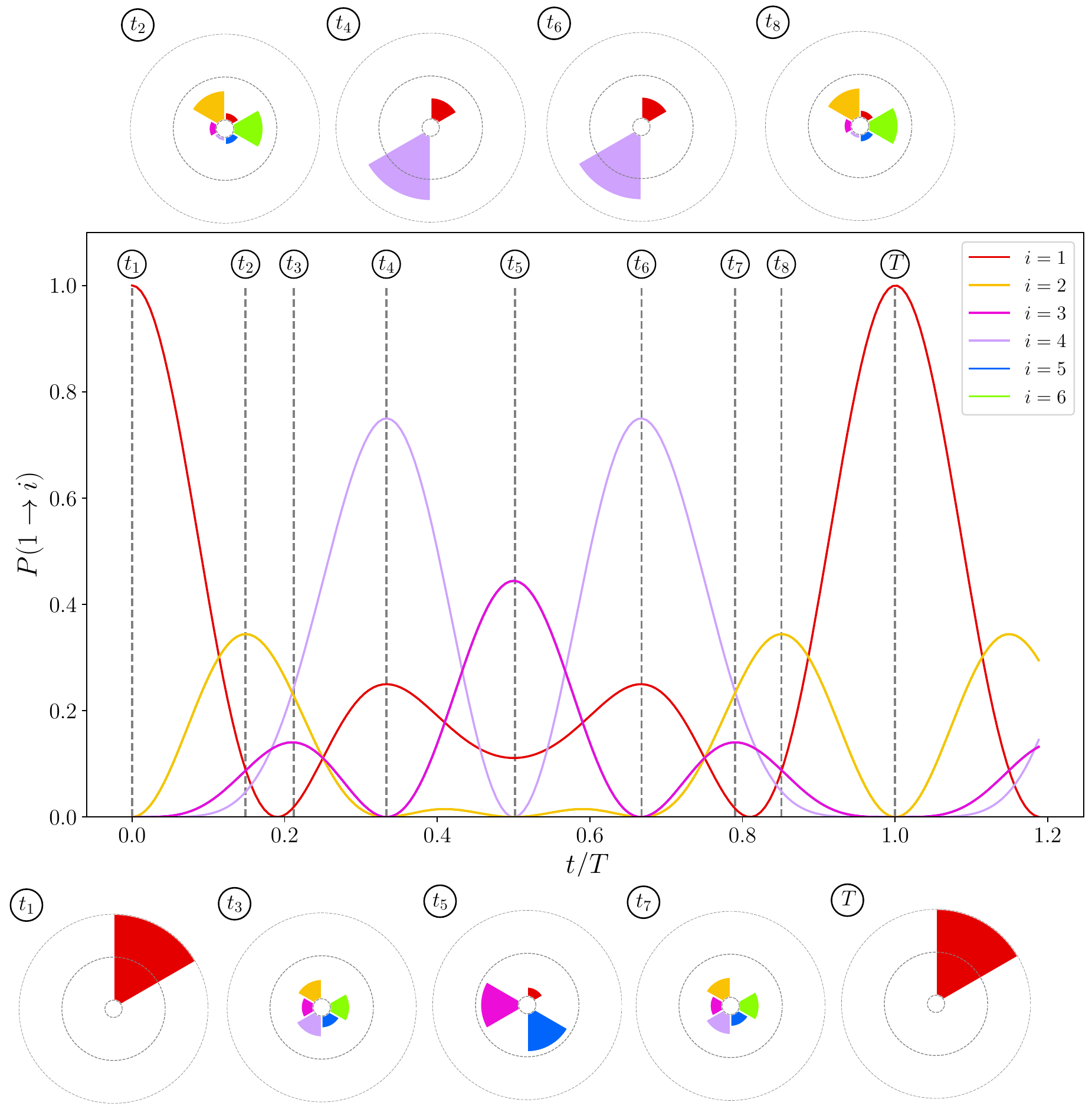}
\caption{The flow of probability around the ring of fermions for $N = 6$ sites and $q = 10^3$. The neutrino is initialised at $t=0$ in the active, $i = 1$ state and, as time evolves, may overlap with the other $N -1$ sites, before returning to a maximum of being detected in the $n =1$ state again after a time $T$. The probability that the system is in the $i^{\textnormal{th}}$ site on measurement after a time $t$ is shown in the main figure, in units of the cycle period $T$. The circular bar charts illustrate, to scale, the flow of probability around the sites, plotted in the same colours as in the main figure, at a number of instances. The inner and outer concentric dashed circles denote measurement probabilities of 0.5 and 1 respectively.}
\label{fig:ths}
\end{figure}

To illustrate this, we consider the specific case of $N=6$, corresponding to one active neutrino site and 5 sterile sites. In the main panel of Fig.~\ref{fig:ths} we consider initialising the system in the active flavour state, which we number as state 1, and plot the overlap of the time-dependent state with the $i^{\textnormal{th}}$ site eigenstate as a function of the measurement time. Due to the discrete rotational symmetry in theory space the curves for $ i = 5$ and $ i = 6$ overlap exactly with those of states $i = 3$ and $i=2$ respectively, so they are not visible on the plot. The inset images show the overlap with each site at a time $t_j$ and thus how the measurement probability is distributed between the $6$ sites.  When viewed chronologically they illustrate the flow of probability in theory-space, which repeats periodically.  As $N$ is increased, the time taken to return to the starting distribution in which the system is in the active state increases, as there are a greater number of sites for the probability to flow through first. 

\subsection{A Phenomenological Ansatz}\label{subsec:Pheno}
The two scenarios considered above are just examples of the rich landscape of BSM possibilities for the neutrino sector. Whilst it is of course possible to construct the relevant spectral function for any given model, calculate the transition probabilities, and extract the bounds from experimental data, this process would need to be undertaken separately for each individual model under consideration.  Given the number and range of theoretical possibilities, a comprehensive survey is not only cumbersome, but fundamentally not feasible. Experimental analyses are thus typically limited to a handful of the simplest models. 

We find that over the energies probed by existing oscillation experiments, the $n$-flavour oscillation probability distributions arising from both the pseudo-Dirac and band models can be sufficiently mimicked by a spectral function consisting of $n$ top-hat functions. With respectively discrete and quasi-continuous spectral functions, the pseudo-Dirac and band cases span a broad landscape of plausible spectral functions and it is thus reasonable to expect the top-hat set-up to be capable of reproducing neutrino oscillation probability distributions for a broad range of microscopic scenarios.   In this section, we will thus derive the general form of the transition probability generated by top-hat functions.

To this end, we begin by focusing on three-neutrino oscillations, replacing the delta functions (three at $m_i^2$) of the canonical spectral function with top-hat functions of (generally different) breadths $b_i$ centred at these values.  Explicitly, we parametrise the spectral function as
\begin{equation}
\label{eq:hat}
 \rho_{ee}(\mu^2) = \left\{ \begin{array}{ll}
          \frac{1}{b_1}|U_{e1}|^2 =  \frac{1}{b_1}  \cos^2 \theta_{12}\cos^2  \theta_{13}  & \qquad,\qquad   m_1^2 - \frac{b_1}{2}  \leq \mu^2 \leq m_1^2 + \frac{b_1}{2}  \\
          \frac{1}{b_2}|U_{e2}|^2  = \frac{1}{b_2}\cos^2\theta_{13}\sin^2\theta_{12}  & \qquad,\qquad   m_2^2 - \frac{b_2}{2}  \leq \mu^2 \leq m_2^2 + \frac{b_2}{2}   \\
              \frac{1}{b_3} |U_{e3}|^2 = \frac{1}{b_3}\sin^2 \theta_{13}  &\qquad,\qquad  m_3^2 - \frac{b_3}{2}  \leq \mu^2 \leq m_3^2 + \frac{b_3}{2} \\
              0  &\qquad,\qquad \mathrm{otherwise} 
        \end{array} \right\} ~~.
\end{equation}
This is illustrated schematically in Fig. \ref{fig:sd}. We note that current measurements of the leptonic mixing matrix indicate $|U_{e3}|^2 \ll |U_{e  1}|^2, |U_{e 2}|^2$.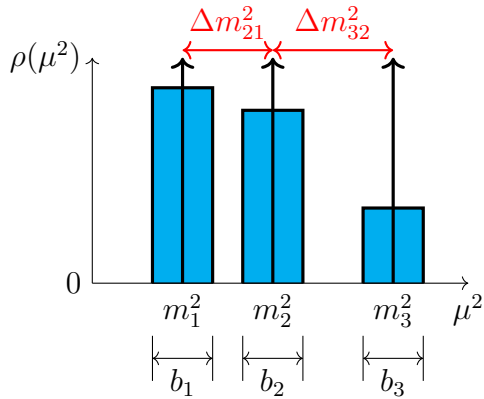
\begin{figure}[t]
\begin{center}
\begin{tikzpicture}

\draw [<->]  (1.6,-1)  -- (0.8,-1);
\draw [<->]  (4.4,-1)  -- (3.6,-1);
\draw [<->]  (2.8,-1)  -- (2,-1);

\draw (0.8,-0.7)  -- (0.8,-1.3);
\draw (1.6,-0.7)  -- (1.6,-1.3);
\draw (2,-0.7)  -- (2,-1.3);
\draw (2.8,-0.7)  -- (2.8,-1.3);
\draw (3.6,-0.7)  -- (3.6,-1.3);
\draw    (4.4,-0.7)  -- (4.4,-1.3);
\draw [fill=cyan] (0.8,0) rectangle (1.6,2.6);
\draw [fill=cyan] (2,0) rectangle (2.8,2.3);
\draw [fill=cyan] (3.6,0) rectangle (4.4,1);
\draw [<-]  (0,3) node [left] {$\rho(\mu^2)$}  -- (0,0);
\draw [<-]  (5,0) node [below] {$\mu^2$} -- (0,0) node[left]{0};
\draw [very thick,black] (0.8,0)  -- (0.8,2.6)  -- (1.6,2.6)
       -- (1.6,0) ;
\draw [ very thick,black] (2,0)  -- (2,2.3)  -- (2.8,2.3)
       -- (2.8,0)  ;
\draw [ very thick,black] (3.6,0)  -- (3.6,1)  -- (4.4,1)
       -- (4.4,0)  ;
\draw [<-,very thick]  (1.2,3)  -- (1.2,0);
\draw [<->,thick,red]  (1.2,3.1)  -- (2.4,3.1);
\draw [<->,thick, red]  (2.4,3.1)  -- (4,3.1);

\node[above,red] at (1.8,3.1) {$\Delta m_{2 1}^2$};
\node[above,red] at (3.2,3.1) {$\Delta m_{3 2}^2$};
\node[below] at (2.4,0) {$m_2^2$};
\node[below] at (1.2,0) {$m_1^2$};
\node[below] at (4.0,0) {$m_3^2$};

\draw [<-,very thick]  (2.4,3)  -- (2.4,0);

\draw [<-,very thick]  (4,3)  -- (4,0);

\node[below] at (1.2,-1) {$b_1$};
\node[below] at (2.4,-1) {$b_2$};
\node[below] at (4,-1) {$b_3$};
\end{tikzpicture}
\end{center}
\caption{Sketch of the form of the  top-hat $\rho_{s}(\mu^2)$ for 3 flavour oscillations as defined in Eq. \ref{eq:hat}. The bold arrows represent the $\delta$-function eigenstates of the standard scenario. }
\label{fig:sd}

\end{figure}

Upon taking the Fourier transform, we arrive at the amplitude
\begin{equation}
iA_{ee} = \sum_{i=1}^{3} \textnormal{sinc}\left(\frac{Lb_i}{4E}\right)|U_{ei}|^2 e^{\frac{-iLm_i^2}{2E}} ~~,
\end{equation} 
and probability  
\begin{align}\label{eq:prob}
P\left(\nu_e \to \nu_e\right) = \left( \sum_i \mathrm{sinc}\left(\alpha_i\right) |U_{ei}|^2\right)^2 - 4\sum_{i<j} |U_{ei}|^2 |U_{ej}|^2 \mathrm{sinc}\left(\alpha_i\right) \mathrm{sinc}\left(\alpha_j\right)\sin^2\left(\Delta_{ji}\right) ~~,
\end{align} 
where $\Delta_{ji} \equiv \Delta m_{ji}^2 L/(4E)$ and $\alpha_{i} \equiv b_i L/(4E)$. We note that since we are treating neutrino propagation in vacuum, the probability for antineutrino oscillations, $\overline\nu_e \to \overline\nu_e$, is identical.

For the case of equal breadths,  $b_1 = b_2 = b_3 = b$, this simplifies to 
\begin{eqnarray}
\label{eq:prop}
P\left(\nu_e \to \nu_e\right) & = & \textnormal{sinc}^2\left(\frac{Lb}{4E}\right)\bigg[1-\sin^2 2\theta_{12} \cos^4 \theta_{13} \sin^2 \Delta_{21} \nonumber \\
& & - \sin^2 2\theta_{13}\left(\cos^2\theta_{12}\sin^2\Delta_{31} + \sin^2\theta_{12}\sin^2\Delta_{32} \right) \bigg] ~~,
\end{eqnarray} 

which we identify as the standard probability expression modulated by a factor of $\textnormal{sinc}^2\left(\frac{Lb}{4E}\right)$. We emphasise that these expressions only depend on the relative spacing of the states in $\mu^2$ and not on their absolute value. 
\begin{figure}[h]
    \centering
   \subfloat[Comparison of 3-flavour oscillation probabilities generated by a density of states comprised of 3 top-hat functions  with fractional breadths $\tilde{b}_i$ = 0.005, and a pseudo-Dirac density of states with masses  $(M_1, M_2, M_3) = (1.23, 1.21, 1.39$) $\times 10^{-4} $eV, selected by performing a least squares fit to the top-hat probability distribution. The lower panel shows the fit residuals.]{\includegraphics[width=0.65\textwidth]{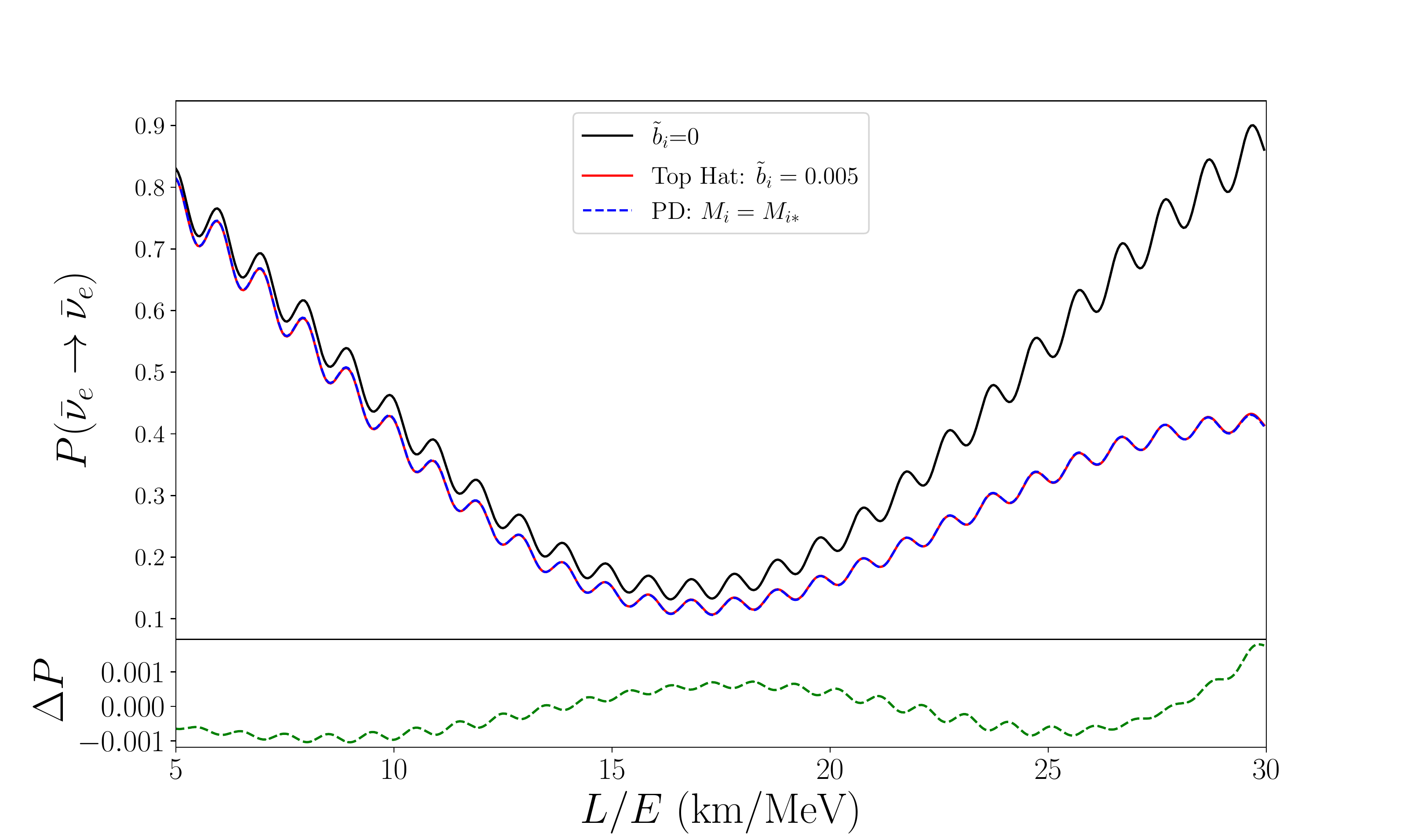} }
   \qquad
    \subfloat[A plot of the spectral function $\rho({\mu^2})$ for these two models with the same parameters as in (a).  Also shown is the triple $\delta$-function density of states corresponding to the canonical scenario. The heights of the top-hat functions relative to each other are plotted to scale but the vertical extent of the (formally infinite)  $\delta$-functions for the pseudo-Dirac and conventional models are intended for illustrative purposes only.]{\includegraphics[width=0.65\textwidth]{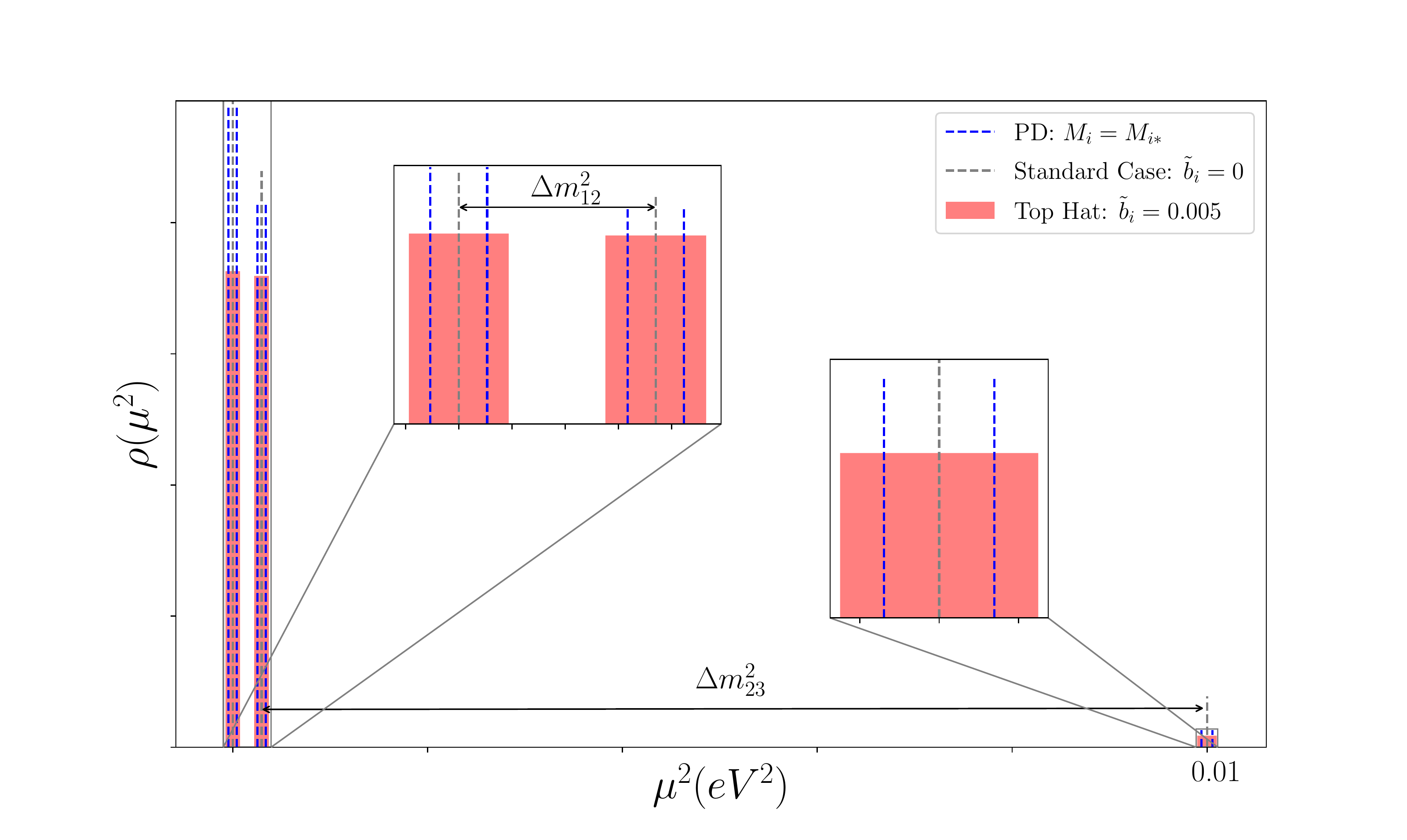} }
\caption{Evaluation of the top-hat phenomenological ansatz defined in Sec.~\ref{subsec:Pheno} to capture the oscillation behaviour of pseudo-Dirac models as detailed in Sec.~\ref{subsec:PD}.}
  \label{fig:0.5pd}%
\end{figure}

\begin{figure}[h]
    \centering
    \subfloat[Comparison of 3-flavour oscillation probabilities generated by a density of states comprised of 3 top-hat functions  with fractional breadths $\tilde{b}_i$ = 0.005, and a $N=10$ band density of states. The value of $q = 987.2$ was selected using a least squares fit to the top-hat probability distribution.]{\includegraphics[width=0.65\textwidth]{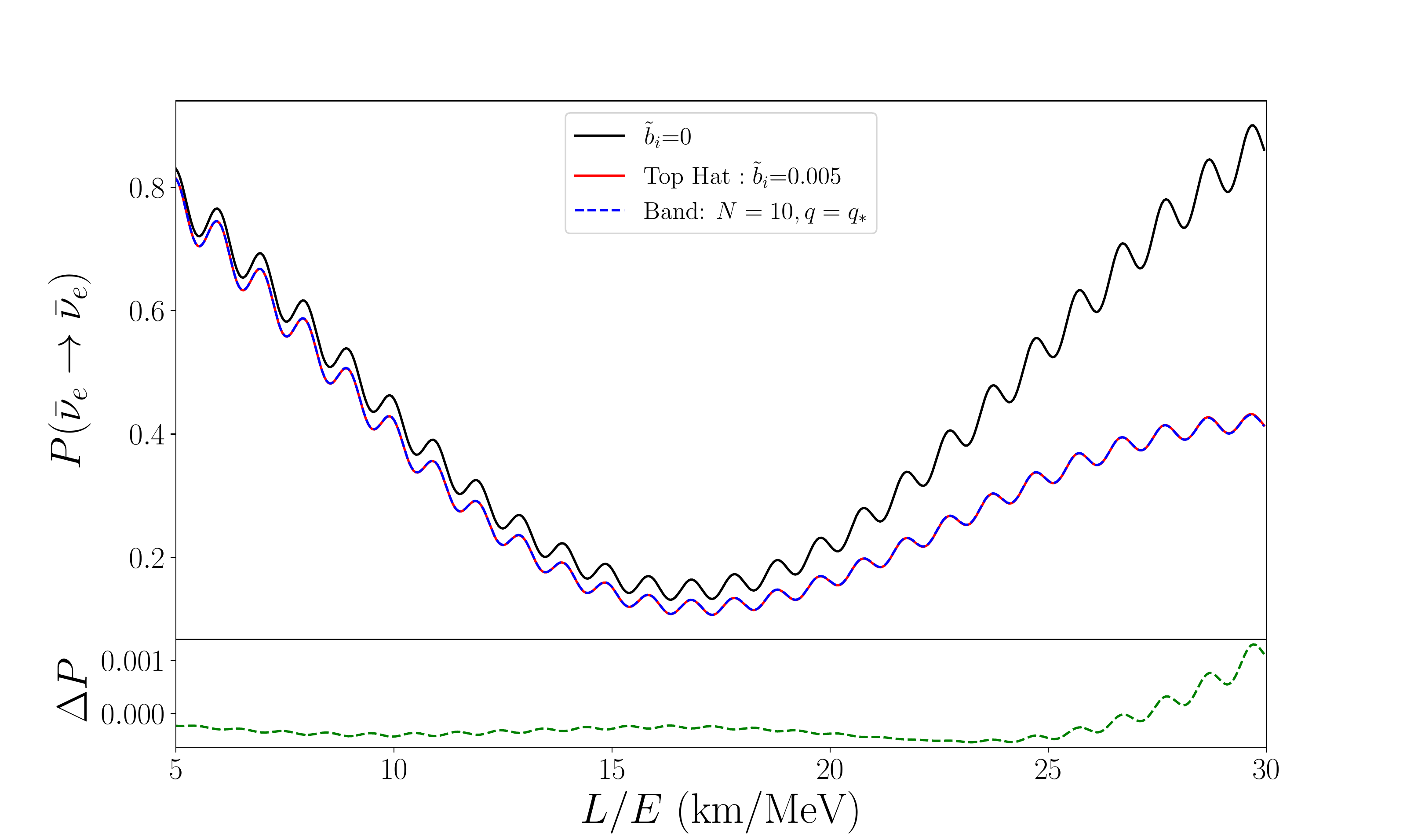} }
    \qquad
    \subfloat[A plot of the spectral function $\rho({\mu^2})$ for these two models with the same parameters as in (a).  Also shown is the triple $\delta$-function density of states corresponding to the canonical scenario. The heights of the top-hat functions relative to each other are plotted to scale but the vertical extent of the (formally infinite)  $\delta$-functions for the band states and conventional model are  intended for illustrative purposes only.]{\includegraphics[width=0.65\textwidth]{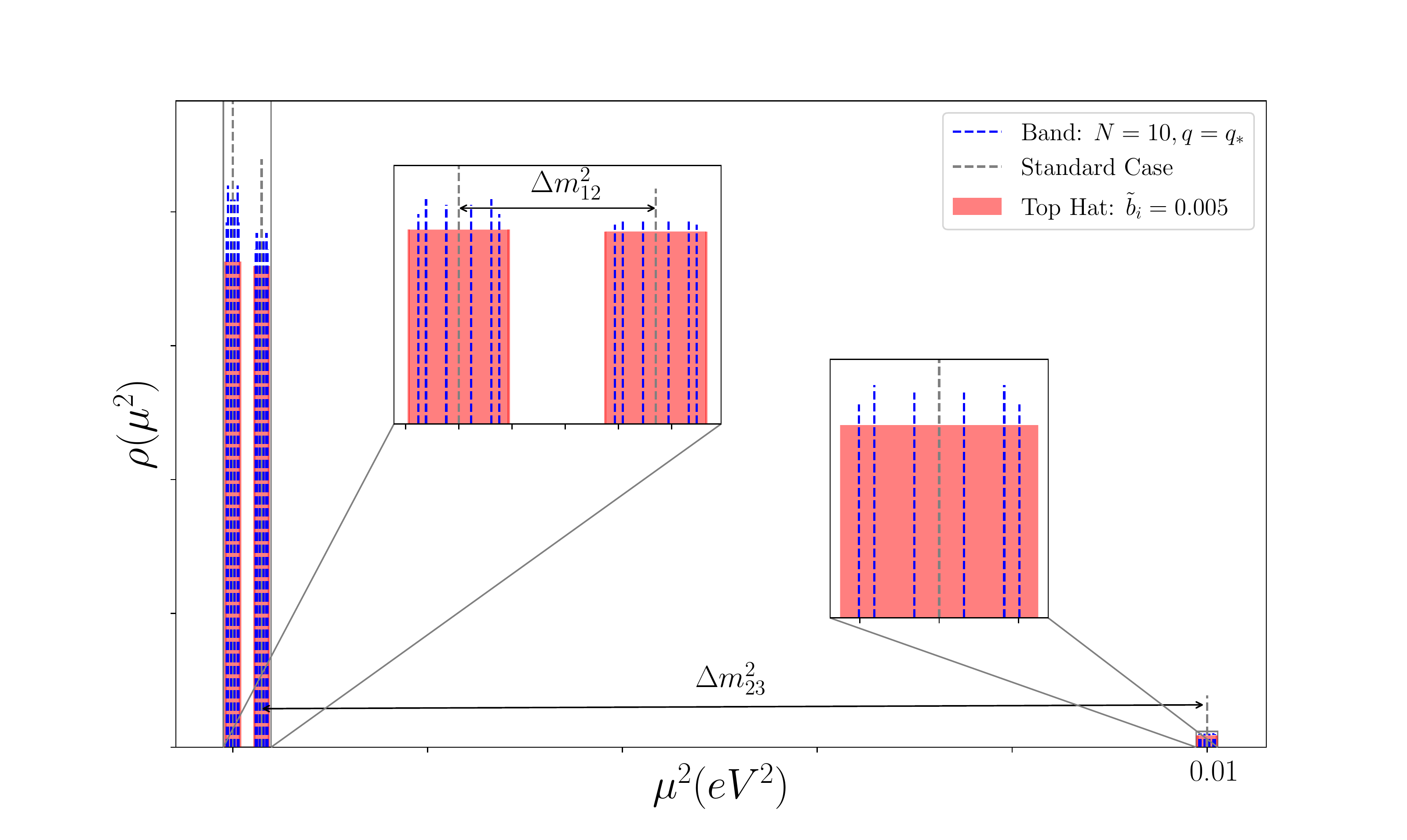} }
    \caption{Evaluation of the top-hat phenomenological ansatz defined in Sec.~\ref{subsec:Pheno} to capture the oscillation behaviour of band models as defined in Sec.~\ref{subsec:Band}.}
        \label{fig:0.5band}%
\end{figure}

For the purpose of this exercise we assume a normal neutrino mass ordering and set  $m_3$ to have an absolute value of 0.1 eV. We then fix $m_1$ and $m_2$ according to $m_i^2 = m_3^2 - \Delta m_{3i}^2$ for $i \in \{1,2\}$. We use the current best fit values of the $\Delta m_{ij}^2$ and the mixing angles as determined by existing experiments\footnote{Explicitly we take $\sin^2\theta_{12} = 0.3$, $\sin^2\theta_{13} = 2.2\times 10^{-2}$, $\Delta m_{21}^2 = 7.4\times 10^{-5}$ eV$^2$, and $\Delta m_{31}^2 = 2.5\times 10^{-3}$ eV$^2$\cite{Esteban:2020cvm}.}.  We parametrise the breadths of the top-hat states according to a fractional value relative to the mass squared, $\tilde{b}_i \equiv b_i/m_i^2$. For reference, we highlight that at $\tilde{b}_1$ = 0.01,  $b_1 \approx \Delta m_{12}^2$ and the spectral gap between the top-hats of state 1 and state 2 vanishes. 

We now seek to demonstrate the capability of the top-hat to capture the phenomenology of the specific microscopic models introduced earlier. To approach this, we generate the (anti-)electron survival probability distribution as a function of baseline length over energy ($L/E$)  for a given set of top-hats breadths $(b_1,b_2,b_3$), and perform a least squares fit of the pseudo-Dirac and band probability distributions over the range $L/E \in [5,30] $ km/MeV,  treating the set ($M_1,M_2,M_3$) as free parameters in the former and fixing $N$ and fitting for $q$ in the latter.  As an illustration, we consider a top-hat set up with fractional breadths $\tilde{b}_i$ = 0.005 which, for reference, corresponds to $b_1 \approx 0.5 \Delta m_{12}^2$. Fig.~\ref{fig:0.5pd} (a) compares the probability distribution for this set up with that generated by the best fit of the pseudo-Dirac model. The best fit parameters are $(M_1,M_2,M_3) = (1.23,1.21,1.39)\times 10^{-4}$ eV. The residuals for the fit are shown in the lower panel and show a disagreement of less than 0.1\% over the domain probed by JUNO. In Fig.~\ref{fig:0.5pd} (b) we plot the density of states for these models with the parameters required for matched probability distributions as detailed above. Also shown is the triple $\delta$-function spectrum of the canonical scenario.  Whilst the heights of the top-hat states relative to each other are plotted to scale, the vertical extent of the (formally infinite) $\delta$-functions for both the pseudo-Dirac model and the standard case are for illustrative purposes only. We observe that probability matching occurs when the breadth of the top-hat function for a given state is roughly twice the pseudo-Dirac mass splitting of that generation. 

In Fig.~\ref{fig:0.5band} (a) we show the agreement between the same top-hat density of states, with the best fit of the band model for $N = 10$, which is achieved for $q = q_* = 987.2$. Once again we see an excellent agreement of within 0.1\% over the  JUNO energy range. As seen in the comparison of the spectral densities for the matched cases shown in Fig.~\ref{fig:0.5band} (b), the breadth of the top-hat of a given state should slightly exceed the breadth of the band. Note that whilst there are 10 mass eigenstates for each generation, degeneracies of the mass values of some states means that they do not appear as distinct lines on the plot. 

Whilst we have merely shown matching to a single choice of top-hat breadths here, we have found that it is always possible to tune the model parameters to achieve an excellent agreement of the probability distributions for any set of top-hat breadths, and thus from the reverse perspective, to be able to find a choice of top-hat breadths which reproduce the probability distributions of these models for any given set of parameters. 

\subsection{The Top-Hat Landscape}
\begin{figure}[t]
\centering
\includegraphics[width=0.75\textwidth]{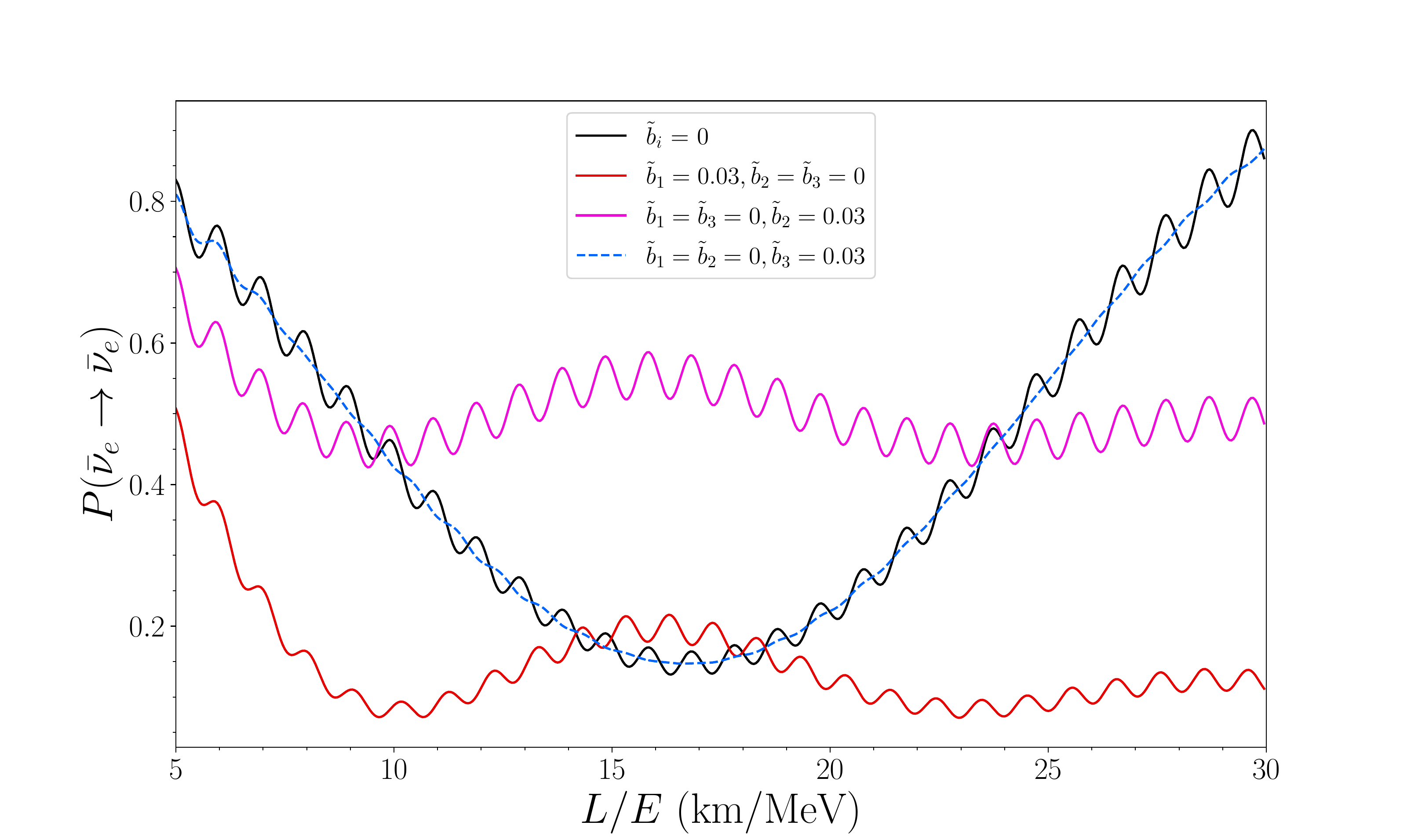}
\caption{Plots of the anti-electron neutrino survival probability as a function of $L/E$ for the 3-flavour top-hat density of states set up as defined in Sec.~\ref{subsec:Pheno} for the case where only one of the three states has a finite breadth, of fractional value $\tilde{b} = 0.03$ and the remaining two are $\delta$-functions. Also shown for comparison is the probability distribution for the standard scenario in which the density of states comprises of 3 $\delta$-functions ($b_i = 0$). }
\label{fig:change1}
\end{figure}

Given their ability to capture the probability distributions of some specific theoretical models, it serves to explore the phenomenological space that can be spanned by top-hat spectral densities in this way. We will address this by exploring the effect on the transition probability of different top-hat breadths. We initially consider the case where only one of the three states has a finite breadth, and the remaining two are $\delta$-functions. In Fig.~\ref{fig:change1} we plot the probability distributions generated by setting  $\tilde{b} = 0.03$ for the non-zero breadth state. This comprises of $\mathcal{O}( 1 )$ amplitude oscillations driven by $\Delta m_{12}^2$ on which a smaller amplitude, higher frequency, oscillation driven by $\Delta m_{13}^2$ is superimposed. We note that the impact of broadening the third top-hat in the spectral function is to damp the amplitude of the $\Delta m_{13}^2$ oscillations. As apparent from Eq.~\ref{eq:prop}, modifications to the large amplitude $\Delta m_{12}^2$ oscillations arise from the sinc terms in $b_1$ and $b_2$, and thus occur on broadening of the breadths of the first and second top-hat states. 

\begin{figure}[h]
\centering
\includegraphics[width=0.75\textwidth]{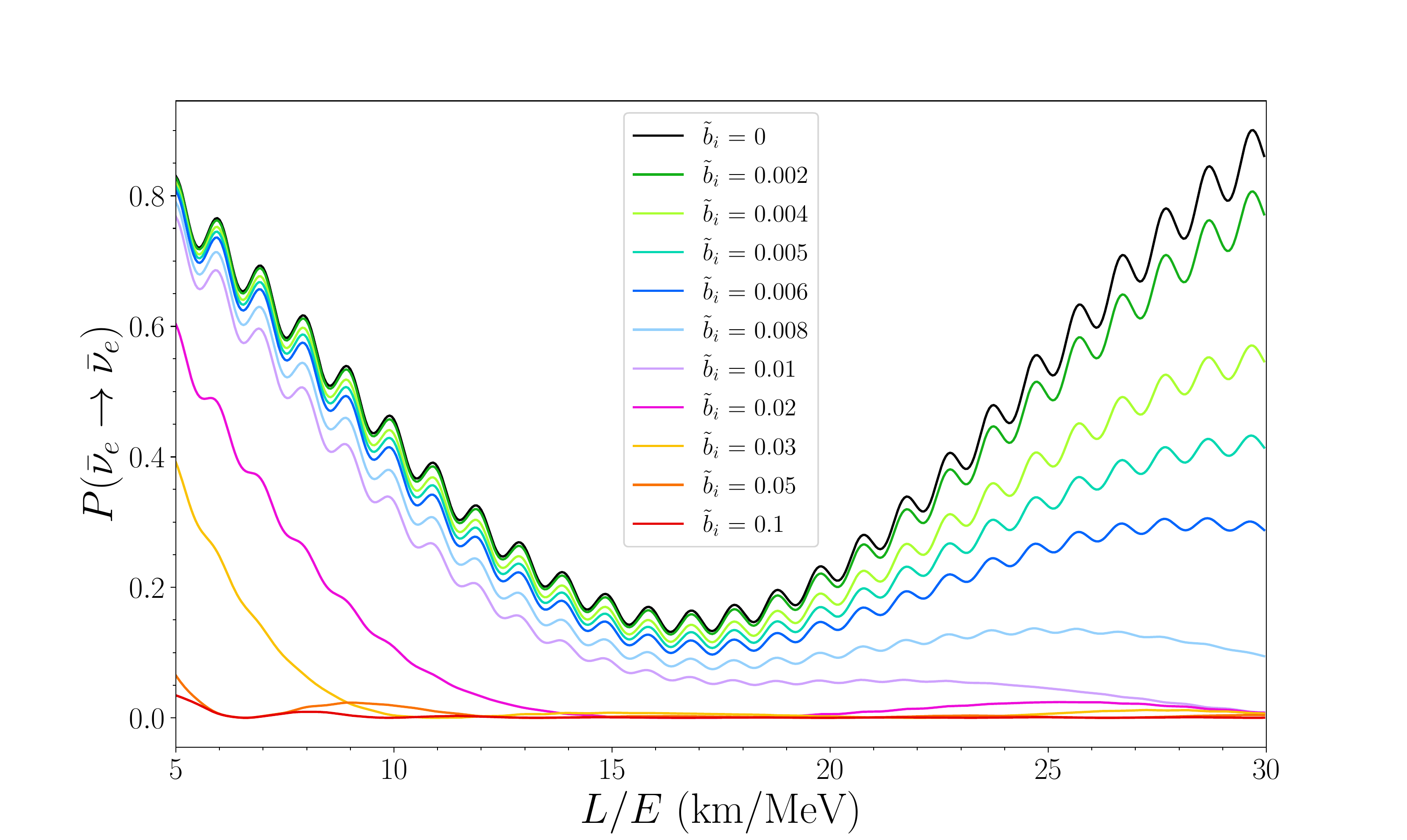}
\caption{Plots of the anti-electron neutrino survival probability as a function of $L/E$ for the 3-flavour top-hat density of states as defined in Sec.~\ref{subsec:Pheno} with  $\tilde{b}_{1}=\tilde{b}_{2}=\tilde{b}_{3}=\tilde{b}$, for various values of $\tilde{b}$. The case $\tilde{b}_i = 0$ corresponds to the standard scenario in which the density of states is comprised of 3 $\delta$-functions. }
\label{fig:comp}
\end{figure}

To gain a handle on the top-hat breadths required to produce measurable deviations from the standard case, we consider the impact on the probability distribution of setting   $\tilde{b}_{1}=\tilde{b}_{2}=\tilde{b}_{3}=\tilde{b}$, such that the breadths of the three states relative to their central mass-squared are equal. We  plot the transition probability for increasing values of $\tilde{b}$ in Fig.~\ref{fig:comp}. Very little deviation from the standard probability distribution occurs for  $\tilde{b}$ $\lesssim$ 0.002, the point at which  $b_1 \approx b_2$ becomes sizeable  ($\sim$ 20\%)  relative to $\Delta m_{12}^2$.  By nature of their larger amplitude, the greatest overall modifications to the probability distribution will arise from modifications to the $\Delta m_{12}^2$ oscillations and thus the degree of deviation from the standard probability distribution is largely controlled by the comparative sizes of $b_1 \approx b_2 $ and  $\Delta m_{12}^2$. Increasing $\tilde{b}_1$ (or $\tilde{b}_2$) both increases the frequency, and decreases the extent of the central peak, of (one of) the sinc functions modulating the $\Delta m_{21}^2$ oscillations.  If $b_1 \ll \Delta m_{21}^2$, the entire energy range probed by JUNO lies approximately at the central peak of the modulating sinc function. In the opposing regime, the JUNO energy range falls in the tails of the sinc function and probability is driven towards zero. If $b_1$ and  $\Delta m_{21}^2$ are of the same order, the probed energy range lies on the falling edge of the sinc function central peak and we see sizeable corrections to the probability, which increase with $L/E$. For reference, we note that for the set of masses used in this figure, $b_1 \approx  \Delta m_{21}^2$ when $\tilde{b}_1 \approx 0.01$ .

\section{The Decoherence Limit}\label{sec:Decoherence}
When propagating over long distances the neutrino wave packets will ultimately decohere, leading to asymptotic neutrino detection probabilities.  One may recall this from the standard two-neutrino treatment, wherein for $\Delta m_{12}^2 L\gg 4E$, there are many oscillations such that \begin{equation}
\sin^2 \left( \frac{\Delta m_{12}^2 L}{4E} \right)  \to \frac{1}{2} ~~,
\end{equation}
and the survival probability asymptotes to
\begin{equation}
P\left(\nu_\alpha \to \nu_\alpha\right) = 1 - \frac{1}{2} \sin^2 2\theta ~~.
\end{equation}
Due to our simplifying ansatz of flavour alignment,  these same limits factorise within the various scenarios considered above, when $L/E$ is much larger than any mass-squared-separation scale of interest for the model at hand.

In this limit, coherence is lost among the mass eigenstates and simplified expressions for the various models may be found.

For the pseudo-Dirac case, assuming all mass eigenstates having splittings such that coherence is lost amongst them, the probabilities become
\begin{eqnarray}
P_{\alpha\beta} & = &  \sum_{i = 1}^{3} (U^*_{\alpha i} U_{\beta i})^2 \left[ \cos^4(\theta_i) + \sin^4(\theta_i) \right] \nonumber \\
& \to & \frac{1}{2} \sum_{i = 1}^{3} (U^*_{\alpha i} U_{\beta i})^2 ~~,
\end{eqnarray}
where in the final expression we have employed the limit $M\ll m_D$, thus $\theta_i \to \pi/4$.  Importantly, note that this is a factor $1/2$ smaller than the corresponding standard three-neutrino result in the decoherence limit.

Now consider the band model, again with a band splitting for each mass eigenstate that is great enough for coherence to be lost amongst the band.  In this case the probabilities become
\begin{eqnarray}
P_{\alpha\beta} & = &  \sum_{i = 1}^{3} (U^*_{\alpha i} U_{\beta i})^2 \sum^{N}_{j= 1}  R_{1 j}^4 \nonumber \\
& \to & \frac{3}{2 N} \sum_{i = 1}^{3} (U^*_{\alpha i} U_{\beta i})^2 ~~.
\end{eqnarray}
The emerging pattern is physically intuitive.  Given a long-enough propagation distance the individual mass wave packets separate.  In these flavour-aligned models, what would have been one mass eigenstate is replaced by $N$ individual separated states, of which an effective $N-1$ are sterile and undetectable.  The overall numerical coefficient depends on the specific details of the model, however the suppression of the resulting signal universally scales inversely proportionally to the number of available states.

Since the top-hat model effectively corresponds to an infinite number of states, one would expect that in the decoherence limit the various detection probabilities would asymptotically vanish.  In this case the electron neutrino survival probability asymptotes to
\begin{eqnarray}
\label{eq:prop}
P\left(\nu_e \to \nu_e\right) & \to & \sum_i \frac{1}{2} \left(\frac{4E}{Lb_i}\right)^2 |U_{ei}|^4 ~~,
\label{eq:thdec}
\end{eqnarray}
which indeed vanishes asymptotically as an inverse quadratic of the length scale.  This will be important when we come to consider experimental constraints.  Furthermore, the mixing angles are also important.  For instance, one has
\begin{equation}
|U_{e1}|^4 \approx 0.5 ~~~~,~~~~ |U_{e2}|^4 \approx 0.09 ~~~~,~~~~ |U_{e3}|^4 \approx 5 \times 10^{-4} ~~~,
\end{equation}
hence, depending on the situation, it may be that constraints on the survival probability of the total electron neutrino flux will only likely significantly constrain broadening in the lightest neutrinos. 

In the following subsections, we discuss how observations of neutrinos from very distant sources can, in principle, place strong constraints on broadened neutrinos through nonzero $b_i$. In contrast, Sections~\ref{sec:Experiments} and~\ref{sec:Constraints} will demonstrate how terrestrial, neutrino-oscillation focused measurements, are an interesting avenue to potentially discover a nonzero $b_i$.

\subsection{Supernova Constraints}
Let us first consider the longest baseline constraints arising from the detection of SN1987A neutrinos by Kamiokande-II \cite{Kamiokande-II:1987idp,Hirata:1988ad}, IMB \cite{Bionta:1987qt,IMB:1988suc}, and Baksan \cite{Alekseev:1988gp}.  Constraints on the pseudo-Dirac case were considered in Ref.~\cite{Martinez-Soler:2021unz}, which informs our comments here.  In Ref.~\cite{Martinez-Soler:2021unz} it was found that constraints were not strong in the decoherence limit since the flux reduction by a factor of 2 could be accommodated by a doubling of the supernova energy.  Stronger limits were found for mass splittings in the $L/E$-range corresponding to SN1987A as this modifies the energy-dependence of the spectrum.  However, in this work we are interested in large mass splittings $\Delta m^2 \gg 10^{-19} \text{ eV}^2$, such that the full decoherence limit is reached.  We would thus expect that in the band model there would be similar flexibility, such that $N=3$ would be acceptable, and perhaps even larger.  On the contrary, for the top-hat ansatz we would expect a significant suppression of the signal for a universal breadth of $b \gg 10^{-19} \text{ eV}^2$.

There is, however, the aforementioned caveat which concerns the universality of the breadth.  We will illustrate this with the top-hat case, although similar aspects apply to the other two models.  Due to the smallness of $|U_{e2}|^4$ and $|U_{e3}|^4$, from Eq.~\ref{eq:thdec} it appears there would be no strong constraint on $b_2$ and $b_3$, since a reduction of the flux due to decoherence in these modes would not be sufficient to generate tension with observations.  On the contrary, we expect a limit in the ballpark of $b_1 \lesssim 10^{-19} \text{ eV}^2$ applies, otherwise the neutrino flux would be too greatly depleted.  The analysis of Ref.~\cite{Martinez-Soler:2021unz} assumed universal splittings, however it would be interesting to repeat this analysis under non-universal assumptions, especially with application to the models considered here.  

\subsection{Solar Constraints}
The case for solar neutrinos is somewhat more complex.  While matter effects are important, it is still ultimately the element $|U_{ei}|^4$ which controls the magnitude of electron neutrino disappearance on broadening the $i^{\textnormal{th}}$ mass eigenstate.  However, the overall fluxes are measured with greater precision and the physics of production understood with greater certainty.  As a result, one again expects the most significant constraints on $b_1$, however constraints on $b_2$ may also be relevant.  Estimating that strong constraints arise whenever $L b_i/E \ll 1$, then for typical solar neutrino energies and the Earth-Sun baseline one expects the limits to be in the region of $b_1 \lesssim 10^{-12} \text{ eV}^2$.

Solar constraints on the pseudo-Dirac case were studied in detail in Ref.~\cite{deGouvea:2009fp}, where indeed limits which would roughly correspond to $b_1 \lesssim 10^{-12} \text{ eV}^2$ were found, however in this case constraints in the region of $b_2 \lesssim 10^{-11} \text{ eV}^2$ also arise, due to the well-measured neutrino flux which gives sensitivity to effects at the $|U_{e2}|^4$-level.  Which width is most strongly constrained is determined from an interplay between the precision of measurement and the magnitude broadening effect which, since it is controlled by the parameter combination $\sim b_j L/E$, is stronger at lower energies for a given fixed width and baseline.

\subsection{Atmospheric \& Astrophysical Constraints}
Due to the typical baseline and energies involved, one expects constraints from atmospheric neutrinos to be at the level of $b_i \lesssim 10^{-4} \text{ eV}^2$.  However, as we will demonstrate in Section~\ref{sec:Experiments}, future long-baseline reactor antineutrino experiments will probe smaller breadths for all $b_i$, and thus we will not consider atmospheric constraints further here.

Finally, the observation of extragalactic neutrinos at neutrino observatories~\cite{IceCube:2013low,IceCube:2014stg} can provide an additional handle on neutrinos traveling great distances. Measuring the ratios of different flavours of the neutrinos upon arrival at Earth~\cite{IceCube:2015gsk,IceCube-Gen2:2020qha} can, in principle, help in constraining many of the models discussed here, however, such constraints would be subject to uncertainties on the overall neutrino flux, among others. Nevertheless, as precision improves~\cite{Song:2020nfh}, the BSM power of these measurements should be considered in more detail.

\subsection{Summary}
It is clear that astrophysical probes of neutrino oscillations allow for baselines that go deep into the decoherence regime.  Indeed, due to the form of the PMNS matrix they lead to very strong constraints on the breadth of the lightest mass eigenstate, at the scale of $b_1 \lesssim 10^{-19} \text{ eV}^2$ and slightly weaker constraints at the level of $b_2 \lesssim 10^{-11} \text{ eV}^2$.  However, they do not probe $b_3$ with the same power due to the smallness of $|U_{e3}|^2$. Moreover, these observations all probe the physics of broadened neutrinos in the decoherence (classical) regime as opposed to making measurements where the propagating neutrinos maintain coherence.
 
As a result, in Section~\ref{sec:Experiments}, we turn to terrestrial probes of neutrino breadths for all mass eigenstates, for their novelty in probing the quantum interference effects of broadened neutrinos, and also as a complementary probe to the methods discussed in this section, subject to a very different set of measurement techniques and assumptions.

\section{Terrestrial Experiments for Constraining Spectral Functions}
\label{sec:Experiments}
Having discussed the strengths and weaknesses of very long-baseline, astrophysical constraints on neutrino breadths in Section~\ref{sec:Decoherence}, we now shift our focus to terrestrial neutrino oscillation experiments. Section~\ref{sec:KLTheory} established our phenomenological description of the neutrino spectral functions; now we explore these experiments and  their ability to test this scenario. In developing this phenomenological approach, we have focused on the case where neutrinos are propagating for long proper times (large $L/E_\nu$) in vacuum, where their interactions with any matter along the path of propagation can be neglected. To date, the experiments consistent with this assumption are those measuring oscillations of electron antineutrinos $\overline\nu_e$ produced in nuclear reactors. These oscillations, with $E_\nu \sim 1{-}10$ MeV, have been measured at a variety of baseline lengths, allowing for world-leading measurements of both mass-squared splittings $\Delta m_{21}^2$ and $\Delta m_{31}^2$ in the standard framework. A second class of experiments measure the oscillations of $\nu_\mu$ produced at ${\sim}$ GeV energies in proton-induced neutrino beams, travelling hundreds of kilometers. These $\nu_\mu$ disappearance experiments offer comparable sensitivity to $\Delta m_{31}^2$ and operate in a similar $L/E_\nu$ regime to the reactor antineutrino experiments.

\paragraph{Reactor Antineutrino $\overline\nu_e$ Experiments}
The most precise measurements of reactor antineutrinos are those from KamLAND~\cite{KamLAND:2008dgz,KamLAND:2010fvi,KamLAND:2013rgu}, with baselines of $L\approx 200$ km, and Daya Bay~\cite{DayaBay:2012fng,DayaBay:2013yxg,DayaBay:2015ayh,DayaBay:2016ggj,DayaBay:2018yms}, with baselines $L \approx 1$ km. These two correspond to measurements at $L/E \approx$ 40 km/MeV and 0.5 km/MeV, respectively. Due to the hierarchicy between $\Delta m_{21}^2 \approx 7.5 \times 10^{-5}$ eV$^2$ and $\Delta m_{31}^2 \approx 2.5 \times 10^{-3}$ eV$^2$, KamLAND is sensitive to oscillations driven by $\Delta m_{21}^2$ (where oscillations due to $\Delta m_{31}^2$ have averaged out over the energy uncertainty of its detector) and Daya Bay is sensitive to oscillations driven by $\Delta m_{31}^2$ (where the oscillations due to $\Delta m_{21}^2$ have yet to develop significantly at the Daya Bay $L/E$ ranges). Because of this, we expect that when studying the generalised spectral functions described in Section~\ref{subsec:Pheno}, KamLAND will be sensitive to nonzero $b_1$ and $b_2$ on the order of $\Delta m_{21}^2$ and that Daya Bay will be sensitive to $b_3$ on the order of $\Delta m_{31}^2$. 

In simulating KamLAND, we develop our analysis to match the results of the most recent collaboration results in Ref.~\cite{KamLAND:2013rgu}. Our simulation of Daya Bay is modified from the analysis of Ref.~\cite{Arguelles:2022bvt} (see Ref.~\cite{Akhmedov:2022bjs} for further discussion), developed to match the official results from Ref.~\cite{DayaBay:2016ggj}.

We also consider the possibility of testing these phenomenological spectral functions in the future, namely by the JUNO~\cite{JUNO:2015zny,JUNO:2022mxj} experiment. JUNO is a medium-baseline reactor experiment that will measure antineutrino oscillations with $L \approx 50$ km, in the $L/E$ regime between that tested by Daya Bay and KamLAND. This will allow JUNO to simultaneously measure oscillations driven by the two mass-squared splittings in a precise way. Previous analyses, including Refs.~\cite{Abrahao:2015rba,Porto-Silva:2020gma,deGouvea:2020hfl,Huber:2021xpx,Basto-Gonzalez:2021aus,Marzec:2022mcz}, have demonstrated that JUNO is an impressive discovery ground for BSM physics -- here, we demonstrate that in our broad-neutrino framework, JUNO will be sensitive to all three spectral-function breadths $b_i$ simultaneously, and should exhibit impressive capability in searching for nonzero $b_i$'s. To simulate JUNO we use the same analysis described in Refs.~\cite{Ellis:2020ehi,Ellis:2020hus}, modified to accommodate our scenario.  We refer the reader to Refs.~\cite{Ellis:2020ehi,Ellis:2020hus} and references therein for more detail.

\begin{figure}[t]
\centering
\includegraphics[width=0.75\textwidth]{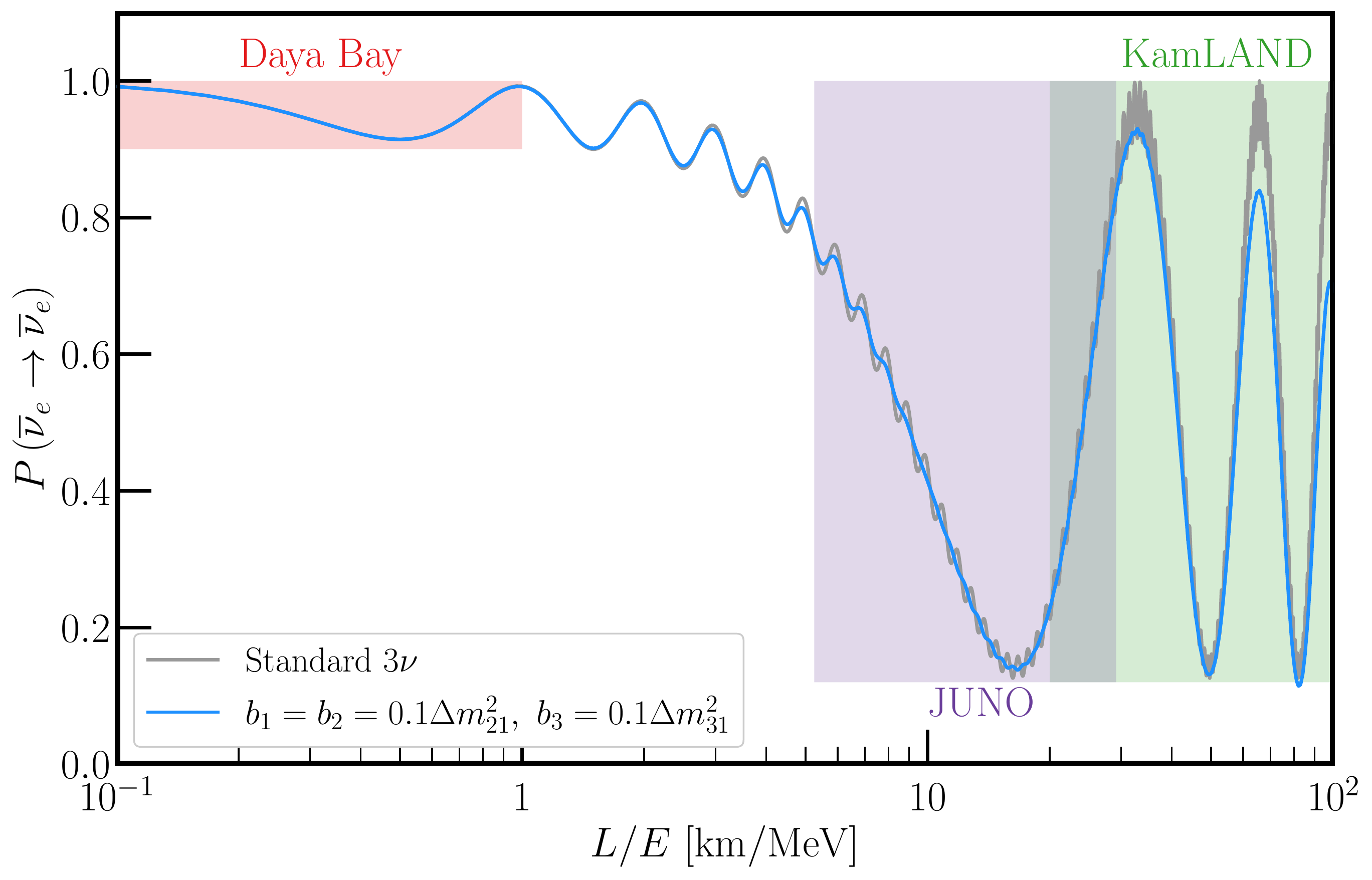}
\caption{Oscillation probability for reactor antineutrinos as a function of $L/E$ for the standard three-neutrino case (grey) and including nonzero spectral-function breadths as indicated in the legend (blue). We shade the regions of $L/E$ probed by existing/future experiments Daya Bay (red), JUNO (purple), and KamLAND (green). \label{fig:PeeLE}}
\end{figure}

The relevant $L/E$ ranges of these three reactor antineutrino experiments are displayed in Fig.~\ref{fig:PeeLE}.
Each experiment's $L/E$ range is shown as a shaded box, with Daya Bay, JUNO, and KamLAND in red, purple, and green, respectively. We also show the oscillation probability $P(\overline\nu_e \to \overline\nu_e)$ as a function of $L/E$ that is/can be measured by these three experiments. We display oscillation probabilities for two cases: grey for the standard three-neutrino scenario\footnote{Assuming the same values for the mass splittings and mixing angles as stated in Sec.~\ref{subsec:Pheno}.} and in blue where we additionally include nonzero spectral-function breadths, $b_1 = b_2 = 0.1\Delta m_{21}^2$ and $b_3 = 0.1 \Delta m_{31}^2$.

Here, the advantage of exploring these effects at large $L/E$ is clear, as in Figs.~\ref{fig:0.5pd}-\ref{fig:comp}. Because KamLAND operated at such large $L/E$, we would expect powerful sensitivity. However, not present in this figure (but present in our simulations) are the effects of finite energy resolution by the respective detectors. For instance, KamLAND has larger energy uncertainty than JUNO will and therefore is not as sensitive to fast, $\Delta m_{31}^2$-driven oscillations in its range of $L/E$ (and does not have sensitivity to $b_3$). Thus, despite its lower $L/E$, we expect JUNO to be the most powerful of these three.

\paragraph{Long-baseline $\nu_\mu$ Disappearance Experiments}
Throughout this work, we are interested in scenarios where neutrino propagation in vacuum is a suitable description. While modern-day (and future) long-baseline experiments measuring $P(\nu_\mu \to \nu_e)$ require the consideration of neutrino interactions with matter for an accurate calculation of oscillation probabilities, the disappearance probability $P(\nu_\mu \to \nu_\mu)$ and its CP-conjugate are insensitive to standard matter effects for the baselines/energies of interest.

For that reason, and for complete comparison against the tests from reactor antineutrino measurements, we include adapted simulations of the T2K~\cite{T2K:2021xwb} and NOvA~\cite{NOvA:2021nfi} experiments from Ref.~\cite{deGouvea:2022kma} to account for these effects in long-baseline $\nu_\mu \to \nu_\mu$ and $\overline\nu_\mu \to \overline\nu_\mu$ oscillations. The oscillation probabilities follow from Eq.~\eqref{eq:prob} with the substitution $|U_{ei}|^2 \to |U_{\mu i}|^2$. We also consider future long-baseline experiments DUNE~\cite{DUNE:2015lol,DUNE:2020ypp} and Hyper-Kamiokande~\cite{Hyper-KamiokandeProto-:2015xww,Hyper-Kamiokande:2018ofw}, which we will comment on in Section~\ref{sec:Constraints}.

In the context of Fig.~\ref{fig:PeeLE}, all of these experiments are situated at a similar $L/E$ to Daya Bay. Therefore, we expect sensitivity to $b_3$ but not competitive with what JUNO will have to offer in the coming decade, due to JUNO's larger $L/E$ and powerful energy resolution.

\section{Current \& Future Constraints on Spectral Functions}
\label{sec:Constraints}
\begin{figure}[t]
\centering
\includegraphics[width=0.75\textwidth]{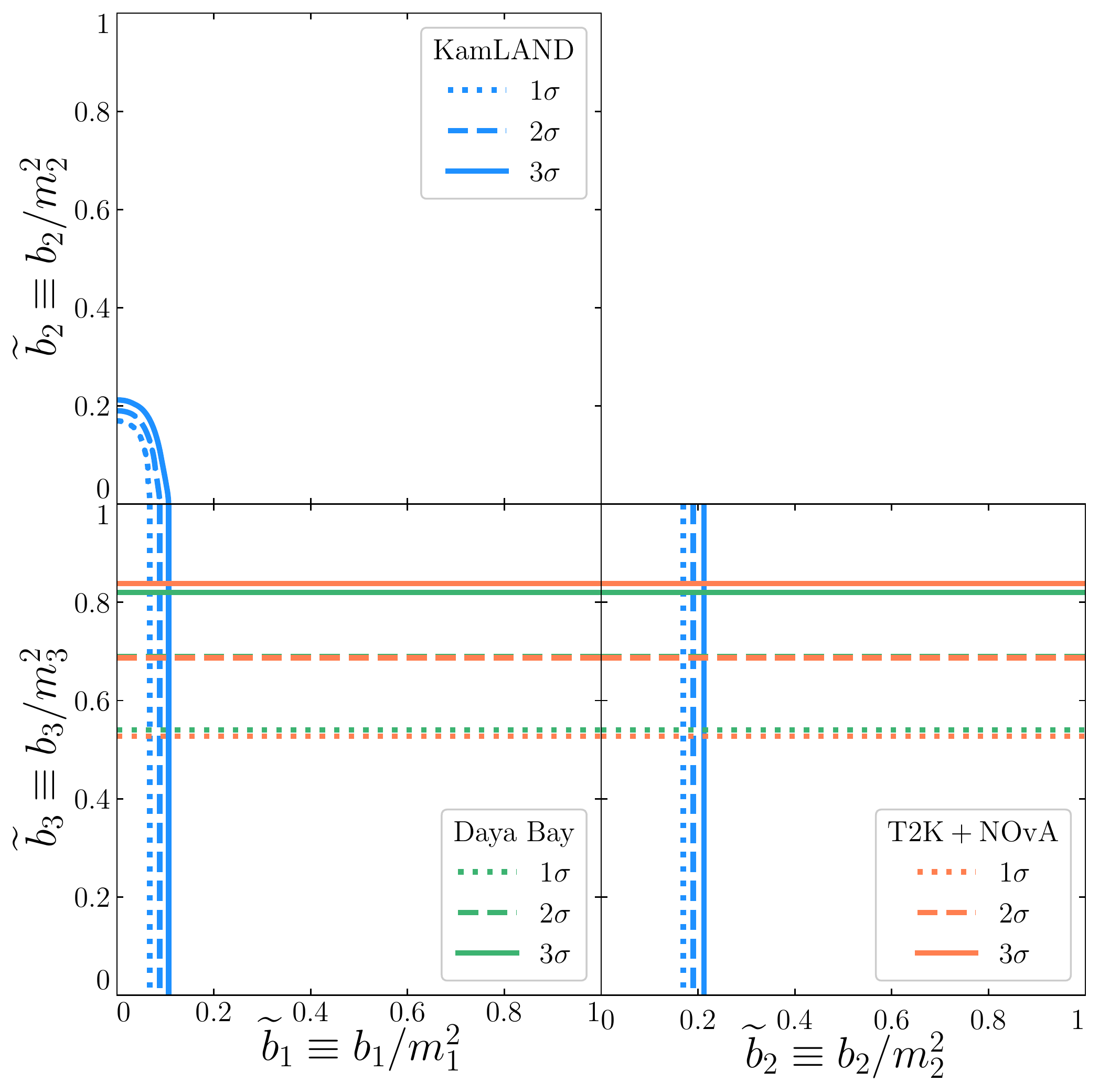}
\caption{Current constraints from KamLAND (blue), Daya Bay (green), and long-baseline $\nu_\mu$ disappearance measurements from T2K and NOvA (orange) at $1$, $2$, and $3\sigma$ CL (dotted, dashed and solid respectively) on the reduced breadths $b_i$, relative to the overall neutrino-masses-squared when we assume $m_1 = 10^{-2}$ eV.\label{fig:CurrentWConstraints}}
\end{figure}
In this section, we provide the current constraints on the breadths $b_1$, $b_2$, and $b_3$. For simplicity, we will focus on the scenario in which the neutrino masses follow the normal ordering $m_3 > m_2 > m_1$. We will present results in terms of the dimensionless $\tilde{b}_i \equiv b_i/m_i^2$. We choose $m_1 = 10^{-2}$ eV as a benchmark for this presentation. Given the discussion in Section~\ref{sec:KLTheory}, we expect that the constraints on $b_i$ from these experiments would be largely unchanged if we considered the inverted mass ordering $m_2 > m_1 > m_3$, however the dimensionless $\tilde{b}_i$ would given the change in the overall $m_i^2$.

When analysing current data, to estimate the constraints on $\tilde{b}_i$, we fix the standard oscillation parameters to their best-fit values as stated in Sec.~\ref{subsec:Pheno}. For our analysis of T2K and NOvA's $\nu_\mu$ disappearance channels, we allow $\Delta m_{31}^2$ and $\sin^2\theta_{23}$ to vary independently.

We present current constraints on the three $\tilde{b}_i$ in Fig.~\ref{fig:CurrentWConstraints}. Notably, we find that current data from KamLAND, Daya Bay, T2K, and NOvA are consistent with $b_1 = b_2 = b_3 = 0$. Here, we compare $1\sigma$ (dotted), $2\sigma$ (dashed), and $3\sigma$ (dashed) constraints for the different experiments, KamLAND in blue, Daya Bay in green, and a combined analysis of T2K and NOvA in orange. As expected from our discussion in Section~\ref{sec:Experiments}, we find that KamLAND has strong sensitivity to $\tilde b_1$ and $\tilde b_2$, but no sensitivity to $\tilde{b}_3$. In contrast, Daya Bay and the long-baseline $\nu_\mu$ disappearance measurements are able to constrain $\tilde b_3$.

\begin{figure}[t]
\centering
\includegraphics[width=0.75\textwidth]{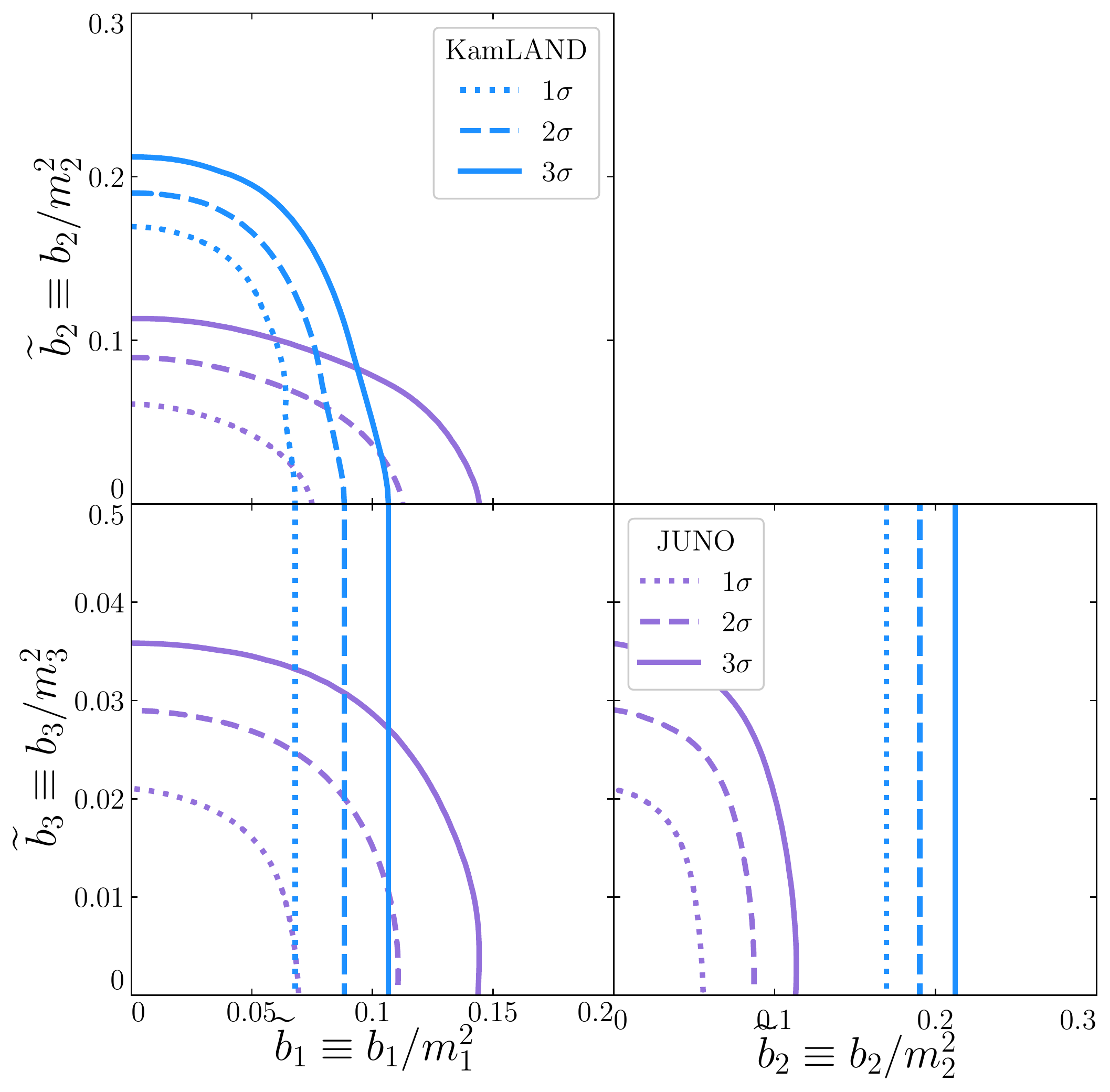}
\caption{Expected future constraints by JUNO (purple) in the absence of a new-physics signal in the parameter space of $\tilde{b}_i$, compared against the current constraints from KamLAND (blue). Compared to Fig.~\ref{fig:CurrentWConstraints}, the data range here is so narrow that the constraints from Daya Bay, T2K, and NOvA do not appear.\label{fig:FutureWConstraints}}
\end{figure}

In contrast, future projections on $\tilde{b}_i$ from JUNO are displayed in Fig.~\ref{fig:FutureWConstraints}, compared against KamLAND's constraints. Note that we have changed the axes ranges here such that the Daya Bay, T2K, and NOvA constraints are no longer visible. Additionally, we have explored the capability of the future Hyper-Kamiokande and DUNE measurements of $\nu_\mu$ disappearance in this context; their sensitivities are both also outside the range shown in Fig.~\ref{fig:FutureWConstraints}.

\begin{table}[h]
\caption{Current and future (expected) terrestrial constraints on $\tilde b_i \equiv b_i/m_i^2$ and $b_i$ at $1\sigma$ confidence.\label{table:UpperLimits}}
\begin{center}
\begin{tabular}{c||c|c||c|c||c|c}
 & $\tilde b_1$ & $b_1$ [eV$^2$] & $\tilde b_2$ & $b_2$ [eV$^2$] & $\tilde b_3$ & $b_3$ [eV$^2$] \\ \hline\hline
Current& $5.7 \times 10^{-2}$ & $5.7\times 10^{-6}$ & $1.6\times 10^{-1}$ & $2.8 \times 10^{-5}$ & $5.3\times 10^{-1}$ & $1.4\times 10^{-3}$ \\ \hline
Future& $5.1\times 10^{-2}$ & $5.1\times 10^{-6}$ & $4.0\times 10^{-2}$ & $7.0 \times 10^{-6}$ & $1.3\times 10^{-2}$ & $3.3\times 10^{-5}$ \\ \hline
\end{tabular}
\end{center}
\end{table}

Taking this set of constraints, we can derive $1\sigma$ upper limits on $\tilde b_i$ as well as $b_i$ given our benchmark $m_1 = 10^{-2}$ eV, which we present in Table~\ref{table:UpperLimits}. While the $\tilde{b}_i$ are useful for dimensionless comparisons, the future sensitivity on the absolute $b_i$ are notable in their own right, demonstrating sensitivity to meV-scale phenomena.

\section{Conclusions}\label{sec:Conclusions}
For decades the neutrino sector has provided us with a unique window through which to study the hidden world of matter. Indeed it remains a theoretically well-motivated location to hunt for new physics. With new neutrino oscillation observatories planned for the near future, it is timely to explore the diverse theoretical and experimental BSM landscape of neutrino oscillations. 

In light of this, we have proposed a new general framework to organise future explorations of the neutrino sector, capturing new physics effects on neutrino propagation in a single spectral function. We demonstrated how this language both reproduces conventional neutrino flavour oscillation calculations, and is simultaneously capable of describing the phenomenology of more exotic theoretical models including those with discrete  and continuous mass spectra. The relevant phenomenological features can in both cases be mimicked by a `toy' mass-spectrum comprising three top-hat functions with a model-specific choice of breadths. We emphasise that this top-hat set up should not be considered a concrete theoretical model, but moreover as a convenient phenomenological ansatz offering the possibility of model-independent analyses of experimental data. Instead of having to sequentially re-interpret searches for different theoretical models, one can equivalently constrain the `breadth' of neutrino spectral functions and thus probe the nature of neutrino propagation effects directly. In the instance of a positive hint for a non-zero neutrino breadth, whilst there may not be a unique invertible mapping from a given set of top-hat breadths to the true mass spectrum, a preference for top-hat functions of non zero breadth is likely to indicate the presence of extra states in the mass-spectrum, and thus the existence of additional sterile neutrinos. 

After discussing how long-distance neutrino measurements can test this non-zero breadth in Section~\ref{sec:Decoherence}, we demonstrated the utility of this approach with terrestrial experiments in Sections~\ref{sec:Constraints}, where we explored the landscape of existing and future oscillation experiments and compared their capacity to probe new physics in neutrino propagation. We found that the long-baseline anti-electron neutrino oscillation experiment KamLAND constrains the breadths of the two lower mass states to remarkable sensitivity but provides no information on the breadth of the third state. Daya Bay and current long-baseline $\nu_\mu$ disappearance searches close this gap somewhat, but current constraints are comparatively weak. The near-future mid-baseline anti-electron neutrino experiment JUNO is projected to significantly improve upon the sensitivities of existing searches, most noticeably for that of the third, highest mass, state. 

We highlight that whilst current data is consistent with the predictions of the conventional 3-neutrino model, the possibility of finite breadths are by no means excluded, particularly for the highest mass state, $m_3$. Finding new physics in the neutrino sector is possible, as testified by the rich and diverse range of BSM scenarios considered in literature.  Armed with the tools to probe the neutrino mass spectrum in a general manner, this framework offers a useful and previously unexploited manner by which to harness the capabilities of future experiments to answer the question:  How broad is a neutrino?

\acknowledgments

We are very grateful to Carlos A. Arg{\'u}elles, Toni Bert\'olez-Mart\'inez, and Jordi Salvado for providing Daya Bay simulation code from Ref.~\cite{Arguelles:2022bvt}, as well as Joachim Kopp for valuable conversations and comments on this draft. HB acknowledges partial support from the STFC Consolidated HEP grants ST/P000681/1 and ST/T000694/1 and thanks members of the Cambridge Pheno Working Group for helpful discussions.

\bibliographystyle{apsrev4-1_title}
\bibliography{biblio}

\end{document}